\documentstyle[12pt]{article}
\begin{document}
\baselineskip21pt
\begin{center} {\bf ROTATING CYLINDRICAL SYSTEMS AND GRAVITOMAGNETISM}
\end{center}

\vspace*{.25cm}

\begin{center} {\bf B. Mashhoon} \\ Department of Physics and Astronomy,
University of Missouri, \\ Columbia, Missouri 65211,
USA \end{center}

\begin{center} {\bf N. O. Santos} \\ Laborat\'orio de Astrof\'{\i}sica e
Radioastronomia \\ Centro Regional Sul de
Pesquisas Espaciais - INPE/MCT \\ Cidade Universit\'aria \\ 97105-900 Santa
Maria RS, Brazil \\ and
\\ Universit\'e Paris VI, CNRS/URA 769,
\\ Cosmologie et Gravitation R\'elativistes \\ Tour 22-12, 4\`eme \'etage,
Bo\^{\i}te 142, 4 place Jussieu, \\ 75005 Paris,
France
\end{center}

\vspace{.25cm}
\centerline{\bf Abstract} We discuss gravitomagnetism in connection with
rotating cylindrical systems. In particular, the
gravitomagnetic clock effect is investigated for the exterior vacuum field
of an infinite rotating cylinder.  The dependence
of the clock effect on the Weyl parameters of the stationary Lewis metric
is determined.  We illustrate our results by means
of the van Stockum spacetime.

\begin{flushleft}PACS numbers: 04.20.Cv; 04.20.Gz\end{flushleft}

\newpage

\section{Introduction} The purpose of this paper is to discuss the physics
of rotating cylindrical systems from the viewpoint
of gravitomagnetism. To this end, we focus attention on the gravitomagnetic
clock effect that exhibits the special temporal
structure around rotating bodies. Briefly, let
$\tau_+$ ($\tau_-$) denote the total proper time of a standard clock that
completes a direct (retrograde) stable circular
geodesic orbit of radius $r\gg 2GM/c^2$ in the equatorial plane of an
astrophysical source of mass $M$ and angular momentum
$J$; then,
\begin{equation}
\tau_+-\tau_-\approx 4\pi\frac{J}{Mc^2},
\end{equation} which is independent of the orbital radius $r$ and the
Newtonian gravitational constant
$G$.  This {\it classical} effect is in some ways reminiscent of the
Aharonov-Bohm effect that is topological in character
[1-3].

It is interesting to observe that for a classical source with uniform
density, the clock effect is independent of $M$.  To
illustrate this point, consider a spherical body of radius $R$ and constant
density rotating rigidly with frequency
$\Omega_0$.  In this case the moment of inertia $I$ is given by (2/5)
$MR^2$ and with
$J=I\Omega_0$ we obtain from (1)
\begin{equation}
\tau_{+}-\tau_{-}\approx\frac{8\pi}{5}\frac{R^2\Omega_{0}}{c^2},
\end{equation}

\noindent which is independent of the total inertial mass of the rotating
body.  Thus the clock effect under consideration
here is a classical gravitational effect that is mainly dependent upon the
{\it rotation} of the source.

Let us now imagine that the shape of the constant density source is changed
to a cylindrical configuration that rotates about
its axis of cylindrical symmetry with frequency
$\Omega_{0}$.  Then equation (1) implies that

\begin{equation}
\tau_{+}-\tau_{-}\approx\frac{2\pi R^2_{0}\Omega_{0}}{c^2},
\end{equation}

\noindent since the moment of inertia of this configuration is
$(1/2)\:MR^{2}_{0}$, where $R_{0}$ is the radius of the circular
section of the cylinder.  These considerations illustrate the dependence of
the clock effect -- which is simply proportional
to the specific angular momentum of the source -- on the shape of the
rotating body.

There are no exact solutions of the gravitational field equations for a
finite rotating cylinder [4].  On the other hand,
equation (3) is independent of the height of the cylinder and this might
suggest that the height of the cylinder is not
relevant and hence we can study the much simpler case of infinite
cylindrical systems; however, the approximation scheme for
which (1) is valid (i.e., $r\gg 2GM/c^2$) applies to the exterior field of
a finite rotating source and would break down for
an infinitely extended cylindrically symmetric configuration of mass-energy
since
$M\rightarrow\infty$.  Nevertheless, we consider in this paper the exterior
gravitational field of infinite rotating
cylindrical systems.  Hence we do not expect to recover equation (3) from
our exact analysis, since our results here apply
only to the ``near zone'' of the rotating cylinder in contrast to previous
work on the gravitomagnetic clock effect that has
been generally concerned with the ``far zone'' of a rotating mass [1-3].
Indeed, the relevance of our results for
cylindrically symmetric systems to naturally occurring (astrophysical)
sources is quite limited.  Instead, it may be possible
to study the influence of certain topological aspects of spacetime
structure -- such as the presence of cosmic strings -- on
the clock effect, since previous work [1-3] has been mainly concerned with
the derivation of the effect in the Kerr geometry
and the possibility of its detection using spaceborne clocks.  However,
only a beginning is made in this first treatment of the
clock effect for cylindrically symmetric fields as we limit our discussion
to the Weyl class of exterior Lewis spacetimes for
the sake of simplicity.

The general relativistic treatment of cylindrically symmetric systems has
been considered by many authors following the early
work of Weyl (cf. [5] for a lucid discussion).  It is interesting to
elucidate certain physical aspects of rotating
cylindrical configurations from the standpoint of gravitoelectromagnetism.
Our treatment here is by no means exhaustive;
instead, we illustrate this point of view by means of an example in this
introductory section and devote the rest of this
paper to the gravitomagnetic clock effect due to its intrinsic physical
significance.

Consider the gravitational field of a ``nonrelativistic'' rotating
astronomical source. In the linear approximation, the
exterior spacetime metric may be expressed as
$g_{\mu\nu}=\eta_{\mu\nu}+h_{\mu\nu}$, where
$\eta_{\mu\nu}$ is the Minkowski metric.  Here Greek indices run from 0 to
3 and Latin indices run from 1 to 3, the
quasi-Cartesian coordinates are
$x^{\mu}=(ct,\vec{x})$ and the signature of the metric is $+2$. Let
$h=$tr$(h_{\mu\nu})$ and define
$\bar{h}_{\mu\nu}=h_{\mu\nu}-\frac{1}{2}\eta_{\mu\nu}h$; then, the
gravitational field equations take the form
\begin{equation}
\Box \bar{h}_{\mu\nu}=-16\frac{\pi G}{c^4}T_{\mu\nu},
\end{equation} once the Lorentz gauge condition
$\bar{h}^{\mu\nu}$$_{,\nu}=0$ is imposed. We are interested in the
particular solution
\begin{equation}
\bar{h}_{\mu\nu}=\frac{4G}{c^4}\int\frac{T_{\mu\nu}(ct-|\vec{x}-\vec{x}^
{\prime}|,
\vec{x}^{\prime})}{|\vec{x}-\vec{x}^{\prime}|}\;\;d^3x^{\prime}.
\end{equation}  For an axisymmetric time-independent source, we find that
the stationary field is given by
$\bar{h}_{00}=4\Phi/c^2$, $\bar{h}_{0i}=-2A_i/c^2$ and
$\bar{h}_{ij}=O(c^{-4})$, where $\Phi(\vec{x})$ is the gravitoelectric
potential,
$\vec{A}(\vec{x})$ is the gravitomagnetic vector potential
$(\nabla\cdot\vec{A}=0)$ and we neglect the tensor potential
$\bar{h}_{ij}$. Thus the exterior metric is of the form
\begin{equation}
ds^2=-\left(1-\frac{2\Phi}{c^2}\right)(d{x^0})^2-\frac{4}{c^4}(\vec{A}\cdot
d\vec{x)}dx^0+
\left(1+\frac{2\Phi}{c^2}\right)\delta_{ij}dx^idx^j,
\end{equation} and we define the gravitoelectric and gravitomagnetic fields by
$\vec{E}=-\nabla\Phi$ and
$\vec{B}=\nabla\times\vec{A}$, respectively, in complete analogy with
electromagnetism.

Imagine now the curvature of spacetime generated by such a source
\begin{equation}
R_{\mu\nu\rho\sigma}=\frac{1}{2}(h_{\mu\sigma,\nu\rho}+h_{\nu\rho,\mu\sigma} -h_
{\nu\sigma,\mu\rho}
-h_{\mu\rho,\nu\sigma}).
\end{equation} The components of the curvature tensor as measured by a
standard set of exterior observers may be expressed as
$R_{\mu\nu\rho\sigma}\lambda^{\mu}_{(\alpha)}\lambda^{\nu}_{(\beta)}
\lambda^ {\rho}_{(\gamma)}\lambda^{\sigma}_{(\delta)}$ and
at the linear order of approximation under consideration here we may set
$\lambda^{\mu}_{(\alpha)}={\delta^{\mu}}_{\alpha}$ either for static
Killing observers or for the free (geodesic) test
observers. The components of the Riemann tensor may be represented in terms
of symmetric and traceless
$3\times 3$ matrices $\cal E$ and $\cal B$ as
\begin{equation}
\cal R=\left( \begin{array}{cc}
\cal E & \cal B \\
\cal B & -\cal E
\end{array} \right).
\end{equation} Here $\cal E$ represents the {\it electric} components of
the curvature tensor,
\begin{equation} {\cal E}_{ij}=-E_{j,i}\;\;\;,
\end{equation} and $\cal B$ represents the {\it magnetic} components of the
curvature tensor,
\begin{equation} {\cal B}_{ij}=-B_{j,i}\;\;\;.
\end{equation} The stationary exterior ``Maxwell'' equations for $\vec{E}$
and $\vec{B}$ are equivalent to the equations that
result from the fact that $\cal E$ and $\cal B$ as given by (9) and (10)
are indeed symmetric and traceless. If the
gravitoelectromagnetic field is uniform, then (9) and (10) imply that
${\cal R}=0$ and the spacetime is flat at the level of
approximation under consideration here.

Let us now consider the gravitational field of a slowly rotating long
cylindrical shell using our approximate treatment. The
analogy with electrodynamics implies that the gravitoelectric field
vanishes inside the shell. On the other hand, the
gravitomagnetic field is uniform inside the shell and vanishes outside.
Hence the spacetime is {\it flat} inside the shell
$({\cal R}=0)$ and the exterior field is {\it static}
$(\vec{B}=0)$. These conclusions agree with the results of the exact theory
[6-8], though the general situation is more
complicated [6].

The basic notions of gravitoelectromagnetism can thus be used to interpret
physically some of the results of the study of
cylindrical systems.  In the rest of this paper, however, we consider
standard clocks on stable circular geodesic orbits
around infinite rotating cylindrical systems and we shall set $G=1$ and
$c=1$ for the sake of simplicity.  The general
exterior vacuum solution in this case is discussed in section 2.  We
confine our discussion throughout to the Weyl class of
exterior stationary spacetimes.  The stable circular geodesic orbits are
studied in section 3.  The clock effect is examined in
section 4, and it is explicitly worked out in the special case of the
exterior van Stockum spacetime in the appendix.  Finally,
section 5 contains a discussion of our main result (49) and a brief
examination of its dependence on the parameters of the
Lewis metric.

\section{Exterior Lewis metric} The exterior spacetime of a rotating
infinitely long cylindrical source has been obtained by
Lewis [9] and is given by
\begin{equation} ds^2=-fdt^2+2kdt\:d\phi+e^{\mu}(dr^2+dz^2)+ld\phi^2.
\end{equation} Here $f(r)$, $k(r)$, $\mu(r)$ and
$l(r)$ are given by
\begin{eqnarray} f=ar^{1-n}-\frac{c_0^2}{n^2a}r^{1+n},\\
k=-abr^{1-n}-\frac{c_0}{na}\left(1-\frac{bc_0}{n}\right)r^{1+n},\\
e^{\mu}=r^{(n^2-1)/2},\\
l=-ab^2r^{1-n}+\frac{1}{a}\left(1-\frac{bc_0}{n}\right)^2{r^{1+n}},
\end{eqnarray} where $n$, $a$, $b$, and $c_{0}$ are constant parameters. We
limit our study here to the Weyl class [10], where
all these parameters are real. If these parameters are complex, then the
solution reduces to the Lewis class [11] that is
beyond the scope of this work. In this paper, we are concerned with the
behavior of certain measurable quantities (``readings
of clocks'') in the Lewis spacetime.  It is therefore important to discuss
the physical significance of the coordinates that
appear in (11).  To this end, let us choose a definite temporal or spatial
scale $\lambda$ and express all spatial and
temporal quantities in units of $\lambda$; then, the spacetime {\it
interval} expressed in terms of such dimensionless
quantities is given by (11)-(15).  In this way, we always deal with
dimensionless quantities insofar as the Lewis metric is
concerned.  Moreover, the nature of the circular cylindrical coordinates in
(11) is intimately connected with the rotating
source, since the general form of the Lewis metric (11) is appropriate for
matching to the interior solution.  We expect that
in the complete absence of matter and radiation, the spacetime would become
Minkowskian; hence, the coordinates in (11) derive
their particular significance in relation to the source.  The exterior van
Stockum solution [12] discussed in theappendix
provides a particularly clear illustration of this circumstance.

The Lewis metric (11) contains a set of dimensionless parameters $(n, a,
b,$ $c_{0})$.  The physical and geometrical
interpretations of these parameters have been the subject of the pioneering
work of van Stockum [12] followed by various
authors (see, e.g., [4-8], [10-11], [13-16] and the references cited
therein).  It follows from these studies that the
parameter $n$ is associated with the Newtonian mass per unit length of a
uniform line-mass source in the low-density
approximation. The parameter $a$ is connected with the constant arbitrary
potential that exists in the corresponding Newtonian
solution. In the static and locally flat limit of the Weyl class, $a>1$
produces a linear energy density along a string. The
parameter $b$ is associated, in the locally flat limit, with the angular
momentum of the spinning string. The parameter
$c_{0}$ is produced by the vorticity of matter when it is represented by a
general stationary completely anisotropic source.
This parameter $c_{0}$ together with $b$ is responsible for the
stationarity of the Weyl-class metric.

It can be proved [10], using the Cartan curvature scalars, that only $n$
helps to curve spacetime locally and that the three
parameters $a$, $b$ and
$c_{0}$ only influence the global structure of spacetime. As a consequence,
the stationary Lewis metric for the Weyl class is
indistinguishable locally from the static Levi-Civita metric [17] as far as
the curvature of spacetime is concerned.

\section{Circular geodesics} To study the geodesics of the Lewis metric
(11), we start with the Lagrangian
\begin{equation} {\cal L}=\frac{1}{2}f\dot{t}^2
-k\dot{t}\dot{\phi}-\frac{1}{2}e^{\mu}(\dot{r}^2+\dot{z}^2)-\frac{1}{2}l\dot
{\phi}^2,
\end{equation} where a dot stands for $d/d\tau$,
$\tau$ being the proper time. From the Euler-Lagrange equations for (16) we
obtain
\begin{eqnarray}
\frac{d}{d\tau}\left(\frac{\partial{\cal L}}{\partial\dot
t}\right)=\frac{\partial{\cal L}} {\partial t}=0,\\
\frac{d}{d\tau}\left(\frac{\partial{\cal
L}}{\partial\dot{\phi}}\right)=\frac{\partial{\cal L}}{\partial\phi}=0,\\
\frac{d}{d\tau}\left(\frac{\partial{\cal
L}}{\partial\dot{z}}\right)=\frac{\partial{\cal L}}{\partial z}=0,\\
\frac{d}{d\tau}\left(\frac{\partial{\cal
L}}{\partial\dot{r}}\right)=\frac{\partial{\cal L}}{\partial r}.
\end{eqnarray} Some aspects of the geodesics of the general Lewis metric
have been studied in [15].  Equations (17)-(19)
reduce to
\begin{eqnarray} f\dot{t}-k\dot{\phi}=E,\\ k\dot{t}+l\dot{\phi}=L,\\
e^{\mu}\dot{z}=P,
\end{eqnarray} where $E$, $L$ and $P$ are constants and represent,
respectively, the total energy, orbital angular momentum
and linear momentum along the
$z$-axis of the test particle divided by its inertial mass. We assume no
motion along the
$z$-axis, hence $P=0$; therefore, we choose $z=0$ and confine our
considerations to the ($x,y$)-plane.  As is well known, the
equation of radial motion (20) has a first integral that is equivalent to
$g_{\mu\nu}\dot{x}^{\mu}\dot{x}^{\nu}=-1$ for the timelike path under
consideration.  This implies that
\begin{equation}
-f\dot{t}^2+2k\dot{t}\dot{\phi}+e^{\mu}\dot{r}^{2}+l\dot{\phi}^{2}+1=0,
\end{equation} so that $\cal{L}$ in (16) is equal to $\frac{1}{2}$ along
the path. From (21) and (22) we have
\begin{eqnarray}
\dot{t}=\frac{El+Lk}{fl+k^2},\\
\dot{\phi}=\frac{-Ek+Lf}{fl+k^2}.
\end{eqnarray} Thus equation (24) can be written as
\begin{equation} e^{\mu}\dot{r}^2+V(r)=0,
\end{equation} where
\begin{equation} V(r)\equiv 1-\frac{1}{fl+k^2}(E^2l+2ELk-L^2f).
\end{equation} Defining new parameters $p$ and
$q$,
\begin{eqnarray}
p\equiv\frac{1}{a}\left[E-\frac{c_{0}}{n}(Eb+L)\right]^2,\\ q\equiv
a(Eb+L)^2,
\end{eqnarray} and using the metric functions, we find that
$fl+k^{2}=r^{2}$ and equation (28) can be represented as
\begin{equation} V(r)=1-\frac{p}{r^{1-n}}+\frac{q}{r^{1+n}}.
\end{equation}

The conditions for circular motion of the test particle at radius
$r=r_0$ are
\begin{eqnarray} V(r_0)=0,\\
\left(\frac{dV}{dr}\right)_{r=r_0}=0.
\end{eqnarray} Furthermore, the circular motion is stable if,
\begin{equation}
\left(\frac{d^2V}{dr^2}\right)_{r=r_0}>0.
\end{equation} The condition (33) gives
\begin{equation} r_0=\left[\frac{(1+n)q}{(1-n)p}\right]^{1/(2n)}.
\end{equation} Only positive roots are to be taken throughout this paper.
Moreover, equations (32) and (35) imply that
\begin{equation} r_0=\left(\frac{2np}{1+n}\right)^{1/(1-n)}.
\end{equation} Hence, $E$ and $L$ can be obtained in principle from (35)
and (36) in terms of $r_0$. From (31) and (35)-(36),
we have

\begin{equation}
\left(\frac{d^2V}{dr^2}\right)_{r=r_0}=\frac{4n^2pq}{r_{0}^4},
\end{equation} which is manifestly positive once
$r_{0}$ can be properly defined.  To satisfy this latter condition, we find
from (35) and (36) that
$-1<n<1$ and $na>0$ must hold.  Then circular geodesic orbits exist and are
stable.

It is interesting to discuss the Newtonian limit of such orbits; to this
end, we set
$f=$exp$(-2\Phi)$ and ignore relativistic rotation parameters $b$ and
$c_{0}$.  Thus to make contact with Newtonian
gravitation via the gravitoelectric potential $\Phi$, we must take $a>0$
and then
$0<n<1$ in this case.  Then, equation (12) implies that

\begin{equation}
\Phi=-\frac{1}{2}{\rm ln}\;a-\frac{1}{2}(1-n){\rm ln}\;r,
\end{equation}

\noindent which should be compared with the corresponding Newtonian
potential for a line-mass, namely,
$-2\sigma$ln $r+$ constant.  It follows from this comparison that
$4\sigma=1-n$, where
$\sigma$ is the mass per unit length of the cylindrical configuration and
the Newtonian limit would then correspond to
$0<\sigma\ll 1$.  We can express $V$ in terms of
$\Phi$ as
\begin{equation} V(r)=1-ap e^{2\Phi}+\frac{q}{ar^2}e^{-2\Phi}
\end{equation}  for $b=c_{0}=0$.  Expanding this expression to first order
in $\sigma$, we recover the correspondence with the
Newtonian theory of gravitation.

Finally, a remark is in order regarding {\it null} circular geodesics of
the spacetime under consideration here.  It is simple
to show by the methods of this section that no such geodesics exist.

\section{Gravitomagnetic clock effect} The aim of this section is to
calculate the proper time difference between two free
test particles, one corotating and the other counterrotating with respect
to the rotating cylindrical source, on an exterior
circular orbit of radius $r$. We do this by using the results of the
previous section on the circular geodesics for the Weyl
class of the Lewis metric and provide a physical explanation for our main
formula describing the clock effect in this case.

In principle, we could integrate around the circle the expression for
$d\tau/d\phi$ that can be obtained from the inverse of
$\dot{\phi}$ given in (26); however, it proves more straightforward to
start from the geodesic equation for radial motion (20)
restricted to the circular orbit.  This implies that
\begin{equation}
f^{\prime}\dot{t}^2-2k^{\prime}\dot{t}\dot{\phi}-l^{\prime}\dot{\phi}^2=0,
\end{equation} where a prime stands for $d/dr$.  Defining the ``angular
velocity'' for a test particle as $\omega={\dot
\phi}/{\dot t}=d \phi / dt$, we obtain from (40) that $\omega =
[-k^{\prime}\pm(k^{\prime
2}+f^{\prime}l^{\prime})^{1/2}]/l^{\prime}$.  From equations (12), (13),
and (15), we find that
$k^{\prime 2}+f^{\prime}l^{\prime}=1-n^{2}$.  Let us now define $N$ and $D$ as
\begin{equation}
N=(1+n)\frac{c_{0}}{na}\left(1-\frac{bc_{0}}{n}\right)r^n+(1-n)abr^{-n}\pm(1-n^2
)^{1/2},
\end{equation}
\begin{equation}
D=\frac{1+n}{a}\left(1-\frac{bc_{0}}{n}\right)^2r^n-(1-n)ab^2r^{-n}.
\end{equation} Then, the angular velocity of the particle is given by
\begin{equation}
\omega=\frac{N}{D}.
\end{equation} If $b=c_{0}=0$, the Weyl-class Lewis metric (12) becomes the
static Levi-Civita metric and (43) reduces to
$\omega=\pm\omega_0$, where $\omega_{0}>0$ is given by
\begin{equation}
\omega_0^2=\left(\frac{1-n}{1+n}\right)\frac{a^2}{r^{2n}}.
\end{equation} With the help of (44), we can rewrite (43) as
\begin{equation}
\omega=\frac{\pm\omega_0+\frac{c_0}{n}}{1-b \left( \pm
\omega_{0}+\frac{c_{0}}{n} \right)}.
\end{equation}
\noindent It is important to note here that $\omega$ vanishes for
$\omega_{0}=\mp c_{0}/n$; therefore, the free particle is in
this case simply static in the spatial coordinates under consideration.
This could come about if the ``centrifugal
repulsion'' balances the gravitational attraction (cf. the appendix).  From
(11) we have for the proper time of circular orbits
\begin{equation} d\tau^2=fdt^2-2kdtd\phi-ld\phi^2,
\end{equation} which becomes
\begin{equation}
\left(\frac{d\tau}{d\phi}\right)^2=f\left(\frac{1}{\omega}+\frac{c_{0}r^{1+n}}{n
af}+b\right)^2-
\frac{r^2}{f}.
\end{equation} We note that
$b+1/\omega=(\pm\omega_{0}+c_{0}/n)^{-1}$ from (45), hence
\begin{equation}
\left|
\frac{d\tau}{d\phi}\right|=\left(\frac{2n}{1+n}\;ar^{1-n}\right)^{1/2}\;\left|\o
mega_{0}\pm\frac{c_{0}}{n}\right|^{-1}.
\end{equation} Integrating this expression, we obtain
\begin{equation}
\tau_{\pm}=
2\pi\left(\frac{2n}{1+n}ar^{1-n}\right)^{1/2}\;\left|\omega_{0}\pm\frac{c_{0}}{n
}\right|^{-1},
\end{equation} where $\tau_+(\tau_-)$ represents the proper time period
registered by a standard clock on a corotating
(counterrotating) circular orbit around the axis of symmetry.  More
generally, the two clocks could both be corotating or
counterrotating as viewed from a rotating system of coordinates (cf. the
appendix).  In any case, in the discussion of the
clock effect we are interested in $\tau_{+}-\tau_{-}$.  Therefore, we
assume in the following that $n \omega_{0} > \left|
c_{0} \right|$ for the sake of concreteness; moreover, the case $n
\omega_{0} \leq \left| c_{0} \right|$ will be discussed
in the appendix for the exterior van Stockum spacetime.  The proper time
difference between the two periods can be obtained
from (49) giving
\begin{equation}
\tau_+-\tau_-=4\pi\left(\frac{2n}{1+n}ar^{1-n}\right)^{1/2}\frac{\frac{c_{0}}{n}
}{
\left(\frac{c_{0}}{n}\right)^2-\omega_0^2}.
\end{equation} It follows from this result that the gravitomagnetic clock
effect is directly proportional to $c_{0}$.  To
understand our main result physically, it is important to develop an
intuitive interpretation of formula (50).  Let us first
note that this result in independent of $b$; therefore, to simplify matters
let us just consider
$b=0$ for the rest of this argument. Hence the physical argument below is
actually valid only for
$b=0$; however, it can be generalized to include $b\neq0$ as discussed in
section 5. In \cite{Silva}, it has been demonstrated
that for $b=0$ the Lewis metric in the Weyl class can be transformed into
the static Levi-Civita metric by making the
coordinate transformation,
\begin{equation}
\bar{\phi}=\phi-\left(c_{0}/n\right) t.
\end{equation} Therefore, starting from this static metric we can recover
the original stationary metric by a rotation of
frequency
\begin{equation}
\Omega=-\frac{c_{0}}{n}.
\end{equation} Following the analysis given in [3], the result (52) means
that for $b=0$ the gravitomagnetic field in the Weyl
class is {\it constant} and just as in the Larmor theorem [18] can be
replaced by a simple Larmor rotation of frequency
$\Omega$.

In the rotating system, consider two clocks moving along the same circular
path with physical radius
$\rho=\rho(r)$ but one going prograde and the other retrograde. The time
$t_{+}$ that takes the clock moving in the prograde
sense to make a complete revolution in the rotating frame is given by
\begin{equation} (v-\rho\Omega)t_+=2\pi\rho,
\end{equation} while the corresponding period
$t_{-}$ for the retrograde orbit is
\begin{equation} (v+\rho\Omega)t_-=2\pi\rho.
\end{equation} Here $v$ is the speed of circular motion in the {\it static}
metric,
\begin{equation} v=\rho\frac{d\bar{\phi}}{dt}=\rho\:\omega_0,
\end{equation} since in this explanation the rotation corresponds to
gravitomagnetism by the gravitational Larmor theorem.
Therefore, from (53) and (54), we have
\begin{equation} t_+-t_-=\frac{4\pi\rho^2\Omega}{v^2-\rho^2\Omega^2},
\end{equation} and with (52) and (55)
\begin{equation}
t_+-t_-=4\pi\frac{\frac{c_0}{n}}{\left(\frac{c_0}{n}\right)^2-\omega^2_0}.
\end{equation} We note that in (57), $\rho(r)$ drops out. In the static
case, the relationship between $t$ and $\tau$ for a
circular orbit is given by
\begin{equation} d\tau^2=ar^{1-n}dt^2-\frac{r^{1+n}}{a}d\bar{\phi}^2,
\end{equation} and considering (44) we have
\begin{equation}
\frac{d\tau}{dt}=\left(\frac{2n}{1+n}ar^{1-n}\right)^{1/2},
\end{equation} if we assume $d\tau/dt>0$. Since (59) is constant for
$r=$constant, we get
\begin{equation}
\tau_+-\tau_-=\left(\frac{2n}{1+n}ar^{1-n}\right)^{1/2}(t_+-t_-).
\end{equation} Putting this expression together with (57), we recover
equation (50).  In this way, the connection between the
main result (50) for the Weyl class of the Lewis metric and the analogue of
the Sagnac effect [19] is established.

Equation (50) and its physical interpretation via the gravitational Larmor
theorem have been presented here explicitly for
$n\omega_{0}>|c_{0}|$; however, our approach is general and can be used to
show that similar results hold in {\it
appropriately modified form} for $n\omega_{0}<|c_{0}|$ as well.

\section{Discussion} The calculation of the gravitomagnetic clock effect
for the exterior field of a rotating cylindrical
configuration of matter has been restricted to the Weyl case, where the
parameters of the general Lewis metric are real; in
fact, a separate analysis is required for the Lewis case, where the
parameters are complex.  The spacetime metric employed
here can be joined to any slowly rotating cylindrical source.  The physical
meaning of the parameters ($n,a,b,c_0$) emerges
from matching the exterior solution to particular interior solutions.  It
follows from such studies that in general $a,b,$ and
$c_0$ are global parameters that could possibly have topological
significance.  The clock effect is meaningful if the static
parameters ($n,a$) are such that $0<n<1$ and
$a>0$.  For instance, the source could be a cosmic string for $n=1$ with a
flat exterior spacetime that would then rule out
circular geodesic orbits for clocks.  Of the stationary parameters
($b,c_0$), only
$c_0$ appears explicitly in the formula for the clock effect.  The prograde
period $\tau_+$ is longer than the retrograde
period $\tau_-$, just as for the exterior equatorial plane of a rotating
mass, if $c_0<0$.  The
clock effect loses its physical significance in the interesting situation where
$c_{0}=\pm n\omega_{0}$ (cf. the appendix).

To gain a physical understanding of the independence of the clock effect
from the stationary parameter $b$, let us introduce a
new {\it periodic} temporal coordinate $\hat{t}$ given by
$\hat{t}=t+b\phi$.  Consider next a rotation of frequency $c_{0}/n$
with respect to the periodic time $\hat{t}$ about the $z$-axis, i.e. $\phi
\rightarrow
\hat{\phi}=\phi-\left(c_{0}/n\right)\hat{t}$.  Under the transformation
$\left(t,
\phi\right)\rightarrow\left(\hat{t},\hat{\phi}\right)$, whose determinant
is unity just like a regular rotation, the Lewis
metric (11) takes the form
\begin{equation}
ds^{2}=-ar^{1-n}d\hat{t}^{2}+e^{\mu}\left(dr^{2}+dz^{2}\right)+\frac{1}{a}r^{1+n
}d\hat{\phi}^{2}.
\end{equation}
\noindent It follows from this result that the physical connection between
the gravitomagnetic clock effect and the analogue
of the Sagnac effect established in section 4 for $b=0$ is valid for
$b\neq0$ as well insofar as time $t$ can be replaced by
{\it periodic} time $\hat{t}$ (with period $2\pi b$).  In fact, equation
(45) implies that
$d\hat{\phi}/d\hat{t}=\pm\omega_{0}$ as would be expected for a circular
orbit in the static spacetime (61).  Indeed, the
metric form (61) appears to be static with respect to $\hat{t}$; however,
since the periodic temporal coordinate $\hat{t}$
does not monotonically increase along future-directed causal curves, Bonnor
[6] has described the spacetime represented by
(61) as only {\it locally} static.  It is important to note that this
global stationarity due to $b\neq0$ does not affect
standard clocks under consideration in this work.  This is ultimately based
on the fact that time is itself measured via a
simple periodic motion.  In our case, for instance, imagine measuring time
by means of a timepiece based on circular motion
with period $2\pi b$.  It should therefore be clear that Bonnor's sound
logical distinction aside, there is no difference
between periodic time and regular time that would be directly measurable by
a clock.  This operational viewpoint is based on
the fact that in timekeeping -- from the simple mechanical timepieces to
modern atomic clocks -- one basically counts a
fundamental period. Upon further elementary transformations
$\tilde{t}=\alpha\hat{t}$,
$\tilde{r}=\alpha^{-1}r$,
$\tilde{\phi}=\hat{\phi}$ and
$\tilde{z}=\alpha^{-1}z$ with $\alpha^{1+n}=a$, the spacetime metric (61)
becomes locally equivalent to Levi-Civita's metric
for a static line-mass with mass per unit length given approximately by
$(1-n)/4$ [6,20].

\vspace*{.5cm}

\noindent {\Large \bf{Appendix:  Clock effect for the van Stockum spacetime}}

\vspace*{.5cm}

The purpose of this appendix is to study equation (49) explicitly for the
rigidly rotating dust cylinder of van Stockum
[12,6].

Imagine a background inertial frame and an infinite cylindrical
configuration of radius $R_{0}$ in which free dust particles
rotate with constant frequency $\Omega_{0}$ about the axis of cylindrical
symmetry.  Let $\beta=R_{0}\Omega_{0}$ be the speed
of the dust at the rim of the cylinder.  Let us now imagine that we
corotate with the cylinder; that is, we choose a comoving
reference frame in which the particles of the dust cylinder are all at
rest.  Moreover, we choose our basic spacetime scale
$\lambda$ to be $\lambda=R_{0}/e^{1/2}$.  Thus the radius of the van
Stockum cylinder in our units is simply $r=e^{1/2}$ and
the frequency of its rotation (relative to the background inertial frame)
is $\alpha=\lambda \Omega_{0}=\beta/e^{1/2}$.  In
the comoving system of coordinates, the interior $\left(0\leq r\leq
e^{1/2}\right)$ van Stockum metric is given by
$$ds^{2}=-\left(dt-\alpha r^{2}d\phi \right)^{2}+e^{-\alpha^{2}r^{2}}
\left(dr^{2}+dz^{2}\right)
+r^{2}d\phi^{2},\eqno(\rm {A}1)$$
which is a solution of the gravitational field equations for dust of density
$\alpha^{2}$exp$\left(\alpha^{2}r^{2}\right)/2\pi$.

The exterior van Stockum spacetime is a member of the Weyl class only for
$0\leq \beta <1/2$.  For $\beta \geq 1/2$, the
exterior solution belongs to the Lewis class.  Thus we limit our discussion
here to $\beta < 1/2$; then, it is
straightforward to show that for the exterior solution $\left(r \geq
e^{1/2}\right)$ in (11)-(15),
$$n=\left(1-4 \beta^{2}\right)^{1/2},\eqno(\rm {A}2)$$
$$a=\frac{1}{2}\left(1+\frac{1}{n}\right) {\rm exp}
\left[-2\beta^{2}/(1+n)\right],\eqno(\rm {A}3)$$
$$b=4e^{1/2}\frac{\beta^{3}}{(1+n)^{2}},\eqno(\rm {A}4)$$
$$c_{0}=-{\alpha}.\eqno (\rm {A}5)$$

Let us first note that from (44), (A2) and (A3) we have $\omega_{0}=\alpha
\chi^{-n}/n$, where $\chi = r/e^{1/2}$.  Thus
$$\pm \omega_{0} + \frac{c_{0}}{n} = -\frac {\alpha}{n} \left( 1 \mp
\chi^{-n} \right),\eqno (\rm{A}6)$$ which
vanishes only for the upper sign at the boundary surface $\chi=1$ and is
negative throughout the exterior region.  Hence no
free particle (``clock'') would stay at rest in the spatial Weyl
coordinates except when it coincides with a dust particle at
the boundary and then $\tau_{+}=\infty$.  Moreover, it is clear from (45)
and (A6) that $\omega < 0$; therefore, both clocks
at a given $r$ are counterrotating as viewed from the comoving frame.  It
then follows from equation (49) that we always have
$\tau_{+} >
\tau_{-}$ in the exterior region just as in the equatorial plane of a
rotating mass [1-3].

The clock effect for the exterior van Stockum spacetime $(\beta < 1/2)$ is
given -- on the basis  of equation (49) -- by
$$\tau_{+}-\tau_{-}=4\pi\frac{n}{\alpha}T_{n}(\chi),\eqno (\rm{A}7)$$ where
$$T_{n}(\chi)=\frac{\chi^{(n+1)/2}}{\chi^{2n}-1}.\eqno (\rm{A}8)$$
The function $T_{n}(\chi)$ diverges at the boundary $\chi=1$ and decreases
monotonically for $1<\chi<\tilde{\chi}$,
where
$\tilde{\chi}^{2n}=(1+n)/(1-n)$.  In fact, $\tilde{r}=e^{1/2} \tilde{\chi}$
is the outer boundary of the region of interest
in this case since the coordinate system is no longer admissible beyond
this point.  It is interesting to note that for $\beta
\ll 1$, $\tilde{r} \simeq \alpha^{-1}$ as would be expected for the radius
of the light cylinder in a frame rotating with
frequency $\alpha$.


\begin{thebibliography}{50}

\bibitem{Cohen}Cohen J M and Mashhoon B 1993 {\it Phys. Lett. A} {\bf 181} 353

\bibitem{Mashhoon}Mashhoon B 1997 {\it Proceedings of the Workshop on the
Scientific  Applications of Clocks in Space} ed L
Maleki (JPL Publication 97-15, NASA) p 41

\bibitem{Mashhoon1}Mashhoon B, Gronwald F, Hehl F W and Theiss D S 1998
{\it preprint} (gr-qc/9804008); Gronwald F, Gruber E,
Lichtenegger H and Puntigam R A 1997 {\it Proc. Alpbach School on
Fundamental Physics in Space} (ESA, SP-420) p 29

\bibitem {Kramer} Kramer D, Stephani H, MacCallum M A H and Herlt E 1980
{\it Exact Solutions of Einstein's Field Equations}
(Cambridge: Cambridge University Press) p 221

\bibitem{Synge}Synge J L 1960 {\it Relativity: The General Theory}
(North-Holland: Amsterdam) ch. VIII

\bibitem{Bonnor}Bonnor W B 1980 {\it J. Phys. A: Math. Gen.} {\bf 13} 2121

\bibitem{Frehland}Frehland E 1971 {\it Commun. Math. Phys.} {\bf 23} 127

\bibitem{Frehland1}Frehland E 1972 {\it Commun. Math. Phys.} {\bf 26} 321

\bibitem{Lewis}Lewis T 1932 {\it Proc. R. Soc. London, Ser. A} {\bf 136} 176

\bibitem{Silva}da Silva M F A, Herrera L, Paiva F M and Santos N O 1995
{\it Gen. Rel. Grav.} {\bf 27} 859

\bibitem{Silva1}da Silva M F A, Herrera L, Paiva F M and Santos N O 1995
{\it Class. Quantum Grav.} {\bf 12} 111

\bibitem{Stockum}van Stockum W J 1937 {\it Proc. R. Soc. Edinburgh} {\bf
57} 135

\bibitem{Bonnor1}Bonnor W B 1992 {\it Gen. Rel. Grav.} {\bf 24} 551

\bibitem{Opher}Opher R, Santos N O and Wang A 1994 {\it J. Math. Phys.}
{\bf 37} 1982

\bibitem{Herrera}Herrera L and Santos N O 1998 {\it J. Math. Phys.,} in press

\bibitem{MacCallum}MacCallum M A H and Santos N O 1998 {\it Class. Quantum
Grav.} {\bf 15} 1627

\bibitem{Levi}Levi-Civita T 1917 {\it Rend. Acc. Lincei} {\bf 26} 307

\bibitem{Mashhoon2}Mashhoon B 1993 {\it Phys. Lett. A} {\bf 173} 347

\bibitem{Stedman}Stedman GE 1997 {\it Rep. Prog. Phys.} {\bf 60} 615

\bibitem{Marder}Marder L 1958 {\it Proc. R. Soc. London A} {\bf 244} 524

\end{thebibliography}
\end{document}